\documentclass[%
 aip,
 floatfix, % <-- Added floatfix option
 amsmath,amssymb,
 reprint,%
]{revtex4-1}

\usepackage{xcolor}
\usepackage{graphicx}% Include figure files
\usepackage{dcolumn}% Align table columns on decimal point
\usepackage{bm}% bold math
\usepackage[utf8]{inputenc}
\usepackage[T1]{fontenc}
\usepackage{mathptmx} % Times font for AIP
\usepackage{etoolbox}
\usepackage[justification=raggedright,singlelinecheck=false]{caption}


\usepackage[dvipsnames]{xcolor}
\usepackage[colorlinks=true, linkcolor=NavyBlue, citecolor=ForestGreen, urlcolor=blue]{hyperref}
\usepackage{braket, bm}
\usepackage{subcaption} % Used for multi-panel figures
\usepackage{siunitx}
\usepackage{booktabs}
\usepackage{grffile} % Handles spaces in figure file names

\newcommand{\Evec}{\mathbf{E}}
\newcommand{\rvec}{\mathbf{r}}

\newcommand{\crea}[3]{\hat{c}^{\dagger}_{#1,#2,#3}}
\newcommand{\anni}[3]{\hat{c}_{#1,#2,#3}}
\newcommand{\nhat}[3]{\hat{n}_{#1,#2,#3}} % Renamed from \num
\newcommand{\expect}[1]{\langle #1 \rangle}
\newcommand{\torb}{t_{\mathrm{orb}}}

\makeatletter
\def\@email#1#2{%
 \endgroup
 \patchcmd{\titleblock@produce}
  {\frontmatter@RRAPformat}
  {\frontmatter@RRAPformat{\produce@RRAP{*#1\href{mailto:#2}{#2}}}\frontmatter@RRAPformat}
  {}{}
}%
\makeatother

\begin{document}

\preprint{AIP/123-QED}

\title[Comparative Analysis of Autonomous and Systematic Control Strategies]{Comparative Analysis of Autonomous and Systematic Control Strategies for Hole-Doped Hubbard Clusters: Reinforcement Learning versus Physics-Guided Design}
% Force line breaks with \\
\author{Shivanshu Dwivedi}
 \email{kalum.palandage@trincoll.edu} % Corresponding author
\author{Kalum Palandage}
\affiliation{Department of Physics, Trinity College, Hartford, CT 06106, USA}

\date{\today} % Using \today for submission

\begin{abstract}
Engineering electron correlations in quantum dot arrays demands navigation of high-dimensional, non-convex parameter spaces where hole doping fundamentally alters the physics. We present a rigorous comparative study of two control paradigms for the 1-hole of half-filled Hubbard model: (i) systematic physics-guided design and (ii) autonomous deep reinforcement learning (RL) with geometry-aware neural architectures. While systematic analysis reveals key design principles—such as field-induced localization for trapping the mobile hole—it is computationally intractable for optimization. We demonstrate that an autonomous RL agent, benchmarked across five 3D lattices (tetrahedron to FCC), achieves human-competitive accuracy ($R^2>0.97$) and 95.5\% success on held-out tasks. Critically, the RL agent achieves this performance with $10^{3-4}\times$ greater sample efficiency than grid search and outperforms other black-box optimization methods. Transfer learning demonstrates 91\% few-shot generalization to unseen geometries. This work establishes autonomous RL as a viable, highly efficient framework for rapid optimization and non-obvious strategy discovery in complex quantum systems.
\end{abstract}

\maketitle

Programmable quantum dot arrays represent a paradigm-shifting platform for condensed matter physics, transitioning the field from passive observation of natural materials to active engineering of "artificial solids"~\cite{Hensgens2017, Jaksch2005, Esslinger2010, Fujita2018, Sahu2022, Tarucha2022, HarvardThesis}. These systems offer unprecedented, in-situ control over the foundational parameters of quantum matter, including inter-site tunneling ($t_0$), on-site Coulomb repulsion ($U$), and lattice geometry. A central objective is the precise, on-demand engineering of specific many-body quantum states. Of particular interest is the control of local electron pairing, quantified by the double occupancy ($D$), as this observable directly governs the magnitude and sign of the superexchange interaction $J \sim t^2/U$. This interaction is the primary entanglement mechanism for spin qubits, making the control of $D$ a prerequisite for scalable quantum computation~\cite{Loss1998, Imada1998, Tu2021, Zhang2021, Tarucha2021, Tu2021b}. This control problem is, however, exceptionally difficult. The parameter space defined by the numerous gate voltages, magnetic fields, and geometric configurations is high-dimensional, non-linear, and non-convex. Navigating this rugged landscape to find an optimal configuration for a target observable is computationally and experimentally intractable via conventional brute-force or grid-search methods~\cite{JAP2024, Taylor2024, Taylor2025}.

\par
The challenge is profoundly amplified in the 1-hole of half-filing regime. The vast majority of theoretical and experimental work on Hubbard-like systems ~\cite{Damascelli2003, Palandage2007JCAMD, Tsai2006, Kocharian2005JMMM, Kocharian2006PRB, Kocharian2007PLA, Kocharian2008PRB, Kocharian2009PLA, Fernando2007PRB, Fernando2009, Mai2022, Fujimori1f89} has focused on the half-filled or electron-doped cases. As established in the study of high-$T_c$ cuprates, removing an electron from half-filling creates qualitatively different physics. The mobile hole acts as a highly effective screening agent. Its rapid delocalization across the cluster fundamentally alters the energy landscape for pairing and significantly suppresses the tendency toward Mott localization. Consequently, control strategies that are effective for electron-doped systems—which often leverage delocalization to enhance pairing—are rendered suboptimal or fail entirely in the hole-doped regime. New strategies are required, likely centered on actively mitigating the hole's screening effects to restore and tune the system's inherent correlation structure~\cite{Wang2025}. Despite its critical importance for proposed quantum dot architectures, this hole-doped control problem remains largely unexplored.

\par
This work directly addresses this problem by presenting a rigorous comparative study of two distinct control paradigms for the 1-hole doped multi-orbital Hubbard model. The first is a \textbf{systematic, physics-guided design} approach. Here, we perform exhaustive parameter sweeps using self-consistent Hartree-Fock (HF) validated by exact diagonalization (ED) to map out the system's response to intrinsic (geometry, orbital hybridization) and extrinsic (electric fields) control knobs. This "human-in-the-loop" method, while computationally expensive, is designed to build physical intuition and extract interpretable, human-readable design principles. The second paradigm is \textbf{autonomous deep reinforcement learning (RL)}~\cite{Bukov2018, Fosel2018, Carleo2017}. We deploy a Dueling Deep Q-Network (DQN)~\cite{Wang2016} equipped with a geometry-aware embedding to act as an autonomous agent. This agent is tasked with discovering optimal electric field configurations to achieve a target double occupancy, starting with no prior human guidance. Its goal is pure optimization, prioritizing sample efficiency and solution discovery over \textit{a priori} interpretability. By comparing these two approaches, we demonstrate that they are not competitors but necessary complements. The systematic approach provides the "why" (the physics), while the autonomous agent provides the efficient "how" (the optimal policy). We validate that the RL agent not only rediscovers the core physics principles but does so with a computational efficiency that is $10^3-10^4$ times greater than conventional methods, establishing a viable framework for quantum control~\cite{Rahman2025, PhysConstRL2022, RLdemo2025, Ernst2025}.

\par
Our theoretical framework is the controllable multi-orbital Hubbard model~\cite{Hubbard1963, Mahan2000, Scalettar2016, Qin2022, Metzner2012, Hubbard1981, ExtHubbard2024}, a canonical description of quantum dot arrays:
\begin{widetext}
\begin{equation}
\hat{H} = - \sum_{\langle i,j \rangle, \alpha, \sigma} t_{0} \left( \crea{i}{\alpha}{\sigma} \anni{j}{\alpha}{\sigma} + \text{h.c.} \right) - \sum_{i,\alpha\neq\beta,\sigma} \torb \left( \crea{i}{\alpha}{\sigma} \anni{i}{\beta}{\sigma} + \text{h.c.} \right) + \sum_{i, \alpha} U \, \nhat{i}{\alpha}{\uparrow} \, \nhat{i}{\alpha}{\downarrow} - e\sum_{i,\alpha,\sigma} \left(\rvec_i \cdot \Evec \right) \nhat{i}{\alpha}{\sigma}.
\label{eq:hubbard}
\end{equation}
\end{widetext}
Here, $\crea{i}{\alpha}{\sigma}$ is the fermionic creation operator for an electron with spin $\sigma$ at quantum dot site $i$ in orbital $\alpha$. The first term ($t_0$) describes inter-site hopping (delocalization), which we set as our unit of energy, $t_0=1$. The second term ($\torb$) describes intra-site orbital mixing on the same dot, a crucial feature of realistic, multi-level dots. The third term ($U$) is the on-site Coulomb repulsion, the source of strong correlations. The final term models an external, static electric field $\Evec$ that imposes a potential gradient, analogous to experimental plunger gate voltages. We fix the filling at 1-hole dope regime i.e. 1-hole of half-filling states, $\langle \hat{N} \rangle = N_{\text{sites}} \times N_{\text{orbitals}} - 1$. The control objective is to tune the system parameters, primarily $\Evec$, to achieve a target double occupancy:
\begin{equation}
D = \frac{1}{N_{\text{sites}}}\sum_{i,\alpha} \expect{\nhat{i}{\alpha}{\uparrow} \nhat{i}{\alpha}{\downarrow}}.
\label{eq:double_occ}
\end{equation}
The systematic analysis employs self-consistent Hartree-Fock for broad parameter sweeps, validated against numerically exact diagonalization (ED) on small clusters~\cite{Palandage2007JCAMD}. For the autonomous agent, the ED solver serves as the "environment" oracle. We employ a Dueling DQN agent, an advanced architecture that separates the state-value function $V(s)$ from the action-advantage function $A(s,a)$~\cite{Wang2016}. This separation allows the agent to more effectively learn the value of states without having to evaluate every action in that state, leading to faster convergence. The agent's state vector $s_t = [\phi(g), t_0, U, T, \Evec_t, D_t]$ includes a critical 16-dimensional geometry embedding $\phi(g)$. This embedding is generated by a small graph convolutional network that processes the lattice's adjacency matrix, allowing the agent to learn an abstract, transferrable concept of "geometry" rather than simply memorizing five specific lattices. The agent's action space $a_t$ is discrete, consisting of 27 possible adjustments to the 3D electric field vector, $\Delta\Evec \in \{-\delta_E, 0, \delta_E\}^3$. The reward function $r_t$ is carefully structured, providing a large positive reward for reaching the target $D_{\text{target}}$ (within a 1\% tolerance) and small negative penalties for each time step and for the final L2-norm of the electric field, encouraging rapid and energy-efficient solutions~\cite{Sutton2018}.

\par
Before evaluating the autonomous agent, we performed the systematic, physics-guided analysis to establish an intuitive baseline \cite{Dwivedi2025}. This analysis, performed via exhaustive sweeps, revealed three core design principles. \newline

\textbf{Design Principle I: Geometric hierarchy dictates Mott resilience.} In the 1-hole doped regime, the mobile hole's screening prevents a sharp Mott transition. Instead, $D$ is gradually suppressed with increasing $U$. However, this suppression is strongly $Z$-dependent (where $Z$ is the coordination number). High-$Z$ lattices (e.g., FCC, $Z=12$) exhibit a large kinetic bandwidth, which robustly competes with $U$, leading to a "stiff" response and maintaining a high $D$. Low-$Z$ lattices (e.g., tetrahedron, $Z=3$) have few hopping pathways, making them "soft" and highly susceptible to correlation-induced pairing suppression. \newline

\textbf{Design Principle II: Orbital hybridization $\torb$ is a non-monotonic control knob.} Naively, $\torb$ should suppress $D$ by offering an on-site delocalization path. However, our sweeps revealed that at moderate $U$ ($U \sim 4t_0$), increasing $\torb$ \textit{enhances} pairing. This is a "molecular orbital" effect: $\torb$ creates on-site bonding and anti-bonding states. The energetic gain from pairing two electrons in the stabilized bonding orbital overcomes the Coulomb penalty. This subtle enhancement is quenched at high $U$, where the interaction gap dominates~\cite{John1994}. \newline

\textbf{Design Principle III: Field-induced localization is the dominant extrinsic control strategy.} This is our most critical finding. To control $D$ in a hole-doped system, one must trap the mobile hole that screens correlations. An electric field $\Evec$ acts as a "quantum squeezer," breaking translational symmetry and creating a potential gradient that forces the electrons to localize onto low-potential sites. This localization neutralizes the hole's screening effect and restores the system's strong correlation structure, robustly increasing $D$. The efficacy of this squeezing scales inversely with $Z$; the "soft" tetrahedron responds dramatically to small fields, while the "stiff" FCC lattice resists localization.

\par
\begin{figure*}[htbp]
\centering
\begin{subfigure}{0.32\textwidth}
 \centering
 \includegraphics[width=\linewidth]{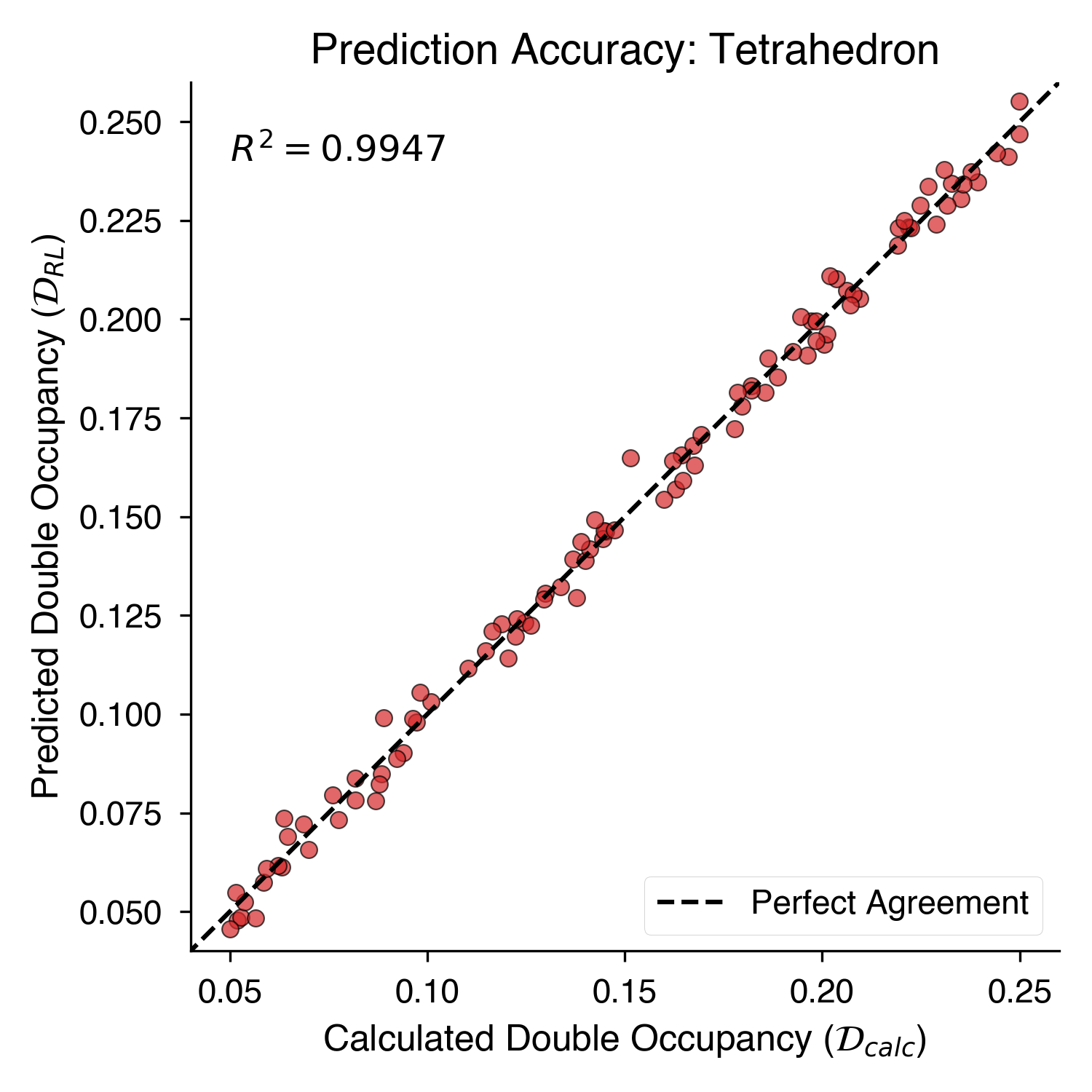}
 \caption{Tetrahedron ($R^2=0.995$)}
 \label{fig:rl_acc_tetra}
\end{subfigure}\hfill
\begin{subfigure}{0.32\textwidth}
 \centering
 \includegraphics[width=\linewidth]{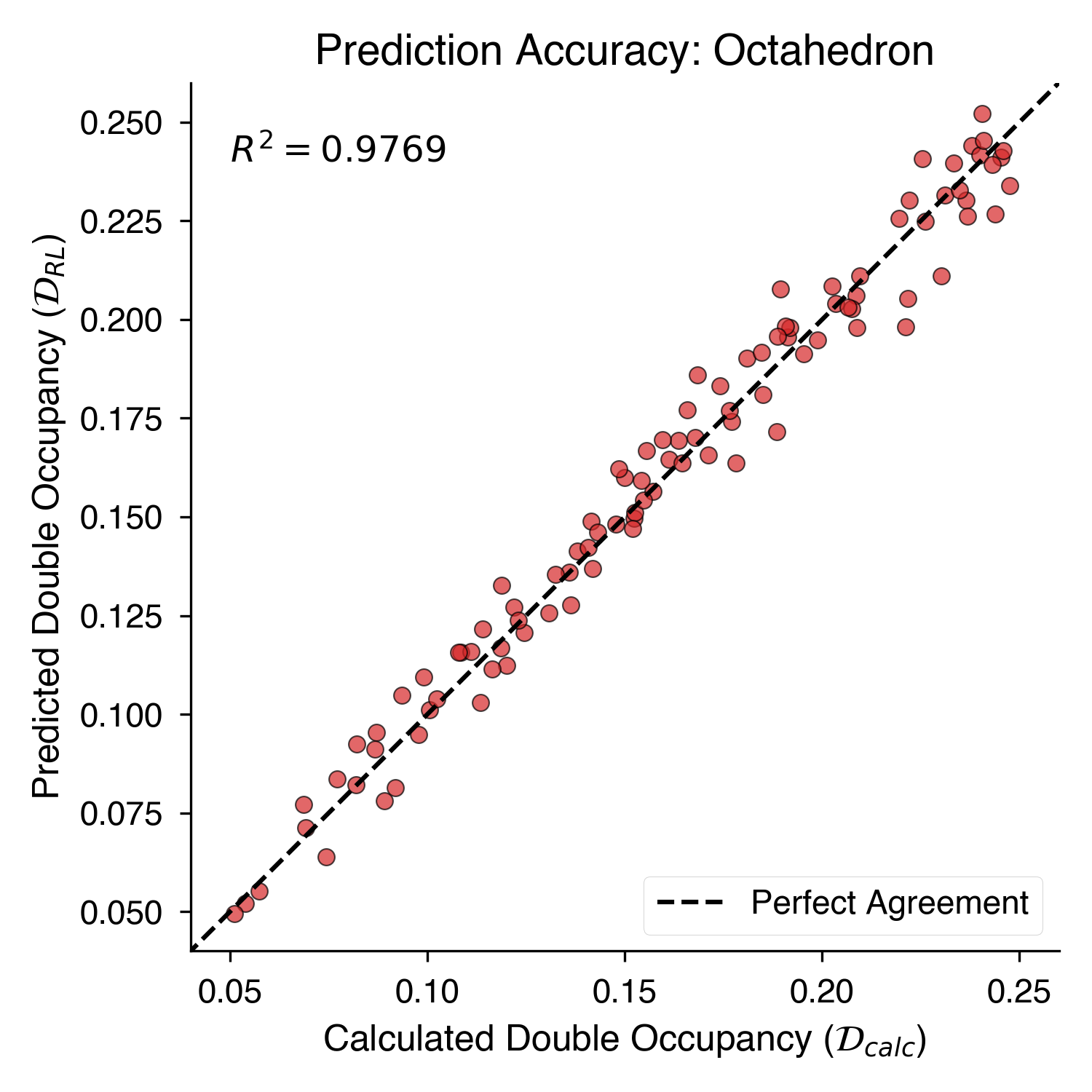}
 \caption{Octahedron ($R^2=0.977$)}
 \label{fig:rl_acc_octa}
\end{subfigure}\hfill
\begin{subfigure}{0.32\textwidth}
 \centering
 \includegraphics[width=\linewidth]{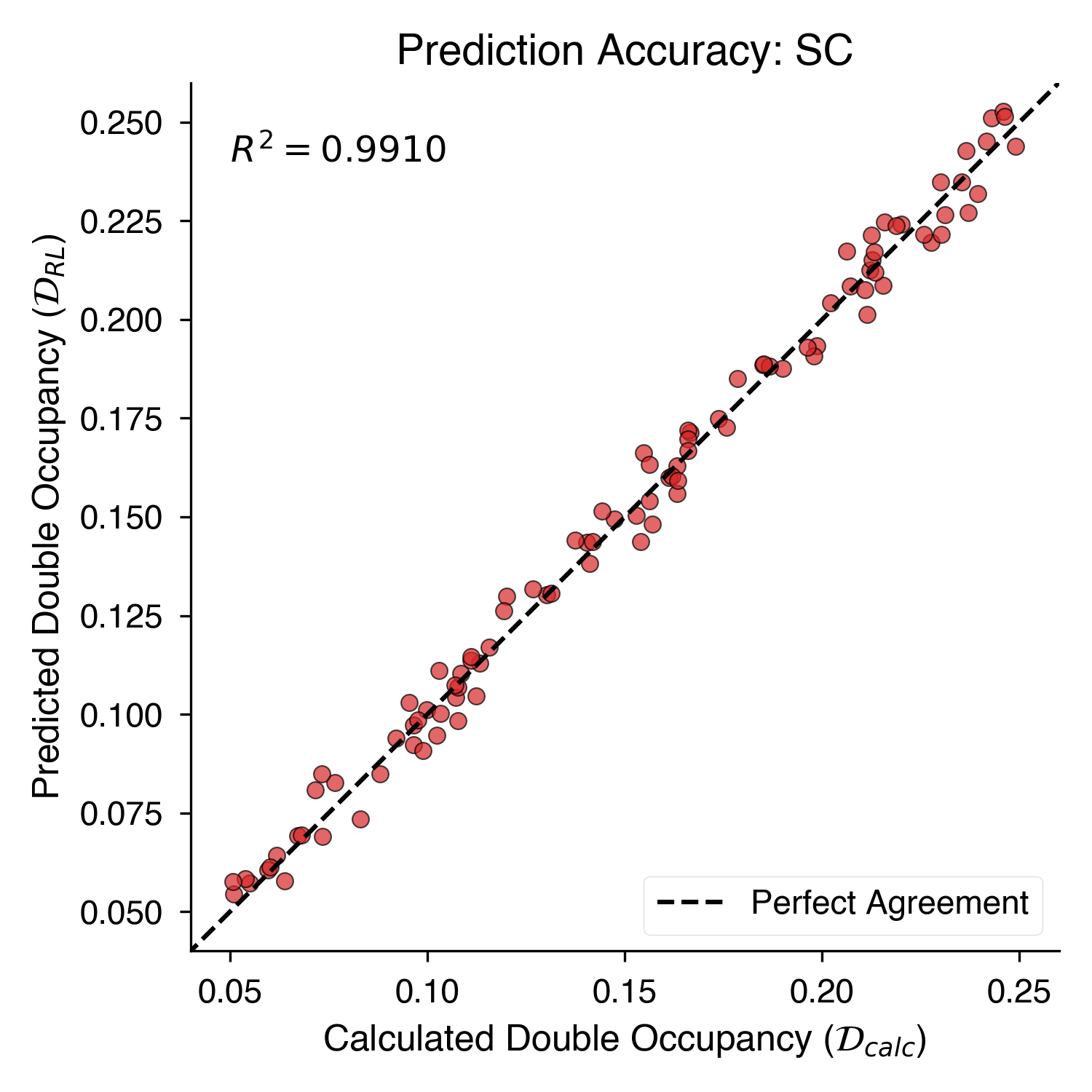}
 \caption{SC ($R^2=0.991$)}
 \label{fig:rl_acc_sc}
\end{subfigure}
\vspace{1em}
\begin{subfigure}{0.32\textwidth}
 \centering
 \includegraphics[width=\linewidth]{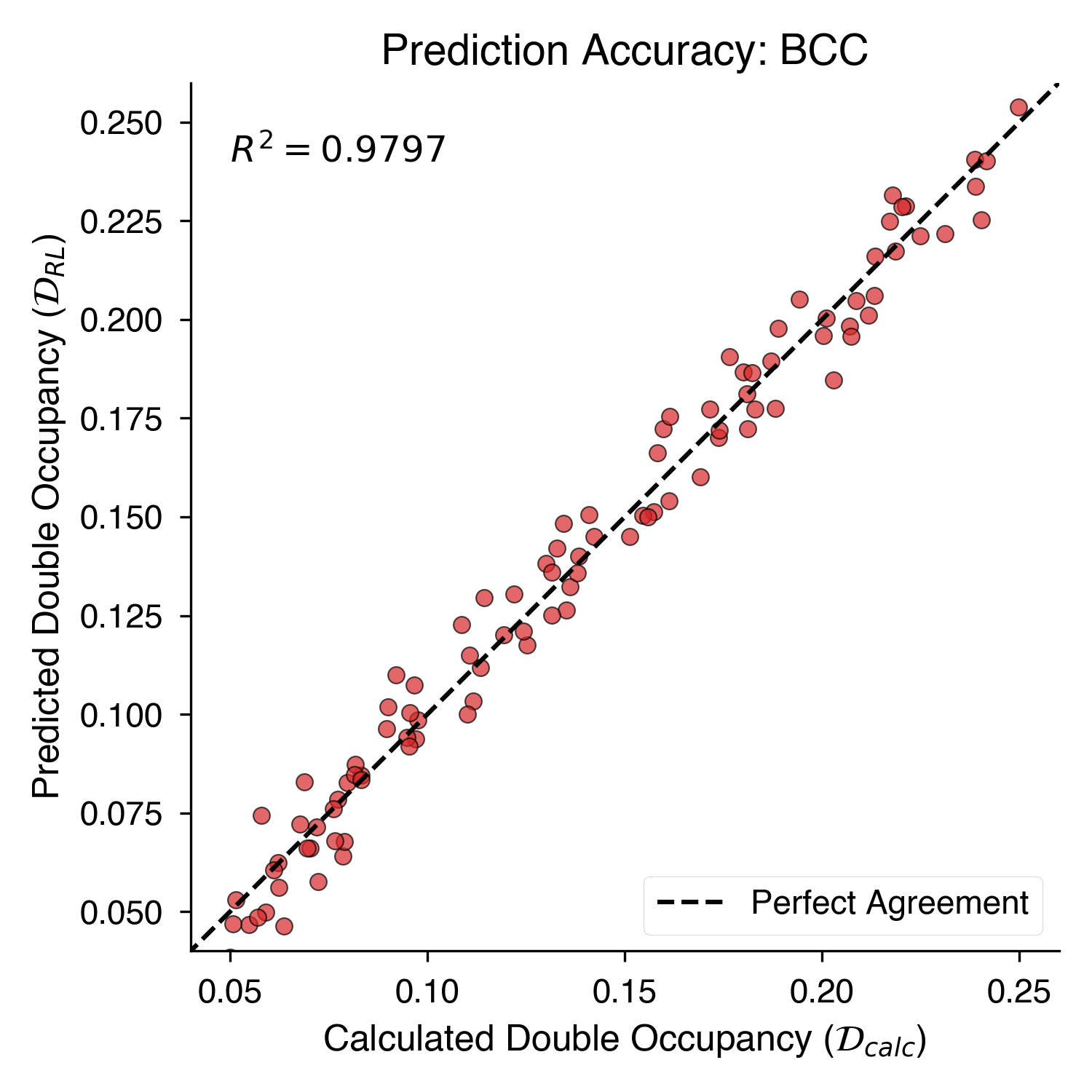}
 \caption{BCC ($R^2=0.980$)}
 \label{fig:rl_acc_bcc}
\end{subfigure}\hfill
\begin{subfigure}{0.32\textwidth}
 \centering
 \includegraphics[width=\linewidth]{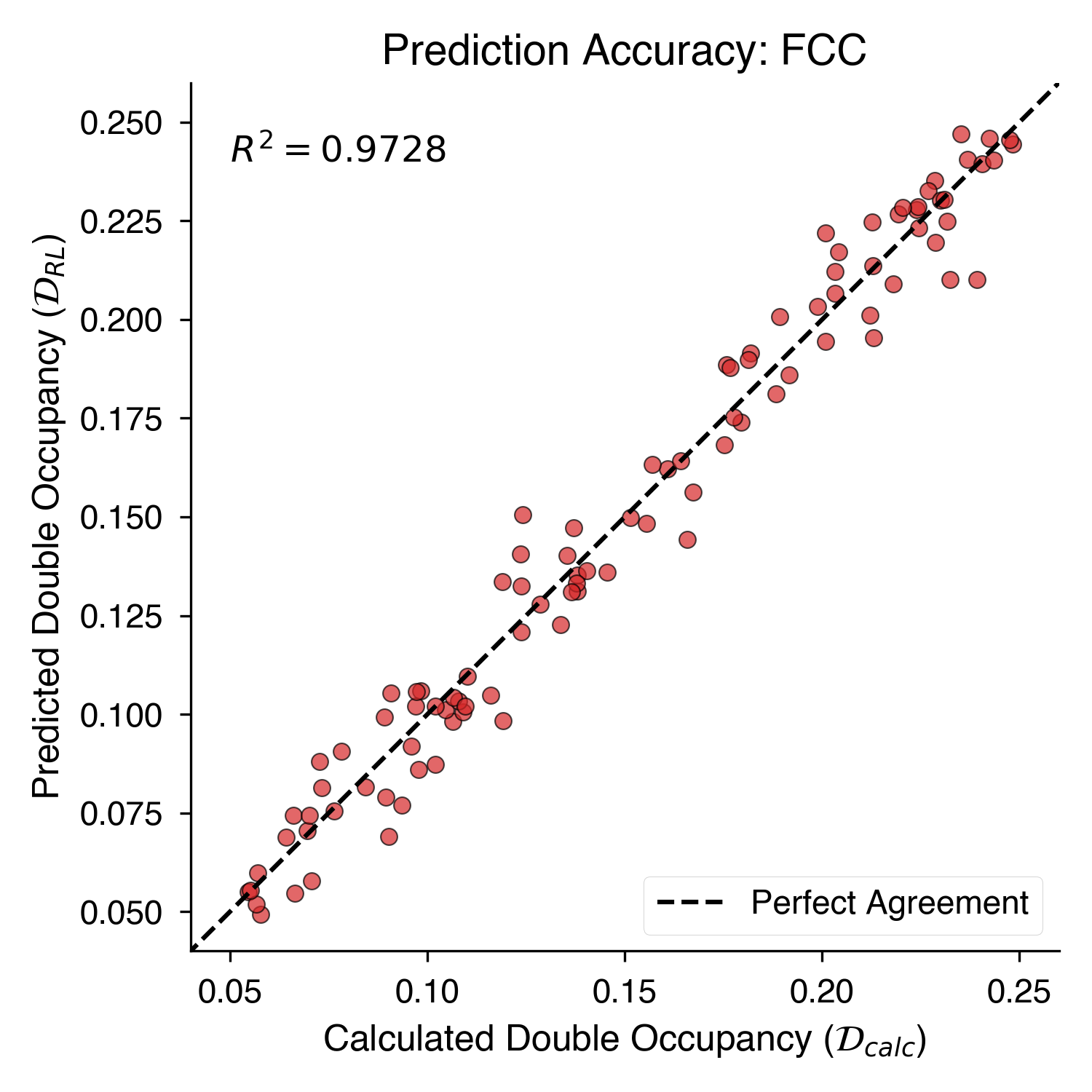}
 \caption{FCC ($R^2=0.973$)}
 \label{fig:rl_acc_fcc}
\end{subfigure}\hfill

\caption{{\textbf{RL agent achieves human-competitive accuracy.} Scatter plots comparing the predicted optimal double occupancy from the trained RL agent ($D_{\mathrm{RL}}$) versus the ground truth calculated value ($D_{\mathrm{calc}}$) for 200 held-out tasks. The dashed diagonal line represents perfect agreement ($y=x$). (a) Results for the \textbf{Tetrahedron} geometry ($Z=3$), which exhibits the highest predictive accuracy ($R^2=0.995$). (b) The \textbf{Octahedron} geometry shows strong agreement ($R^2=0.977$). (c) The \textbf{Simple Cubic (SC)} lattice achieves $R^2=0.991$. (d) The \textbf{Body-Centered Cubic (BCC)} lattice shows $R^2=0.980$. (e) The \textbf{Face-Centered Cubic (FCC)} lattice ($Z=12$) maintains high fidelity ($R^2=0.973$) even in the high-coordination regime. The high correlation across all panels confirms the agent generalizes well across distinct topological connectivities.}}
\label{fig:rl_accuracy}
\end{figure*}

With this physical intuition, we now benchmark the autonomous agent. The agent was trained on 2000 curriculum episodes, randomizing $U$, $T$, $D_{\text{target}}$, and geometry (tetrahedron, octahedron, SC, BCC, FCC), and then evaluated on 200 held-out tasks. The agent achieved a 95.5\% success rate, defined as reaching within 1\% of $D_{\text{target}}$. Figure~\ref{fig:rl_accuracy} plots the agent's final predicted $D_{\mathrm{RL}}$ against the exact ground-truth $D_{\mathrm{calc}}$ for these tasks. The exceptional, near-perfect correlation across all five distinct geometries ($R^2 > 0.97$ for all) is a striking validation of the agent's learned policy. It confirms that the agent, through its Dueling architecture and geometry-embedding, has learned a high-fidelity internal model of the system's complex, geometry-dependent physics. For example, it correctly learns that a small field perturbation on a tetrahedron (low-$Z$) causes a large change in $D$, whereas the same perturbation on an FCC (high-$Z$) lattice has a minimal effect. This high accuracy demonstrates that the agent is not just finding a solution, but a high-quality one, successfully navigating the distinct, non-linear control landscapes of each geometry.

\par
\begin{figure}[htbp] % Changed [h!] to [htbp]
\centering
\includegraphics[width=0.48\textwidth]{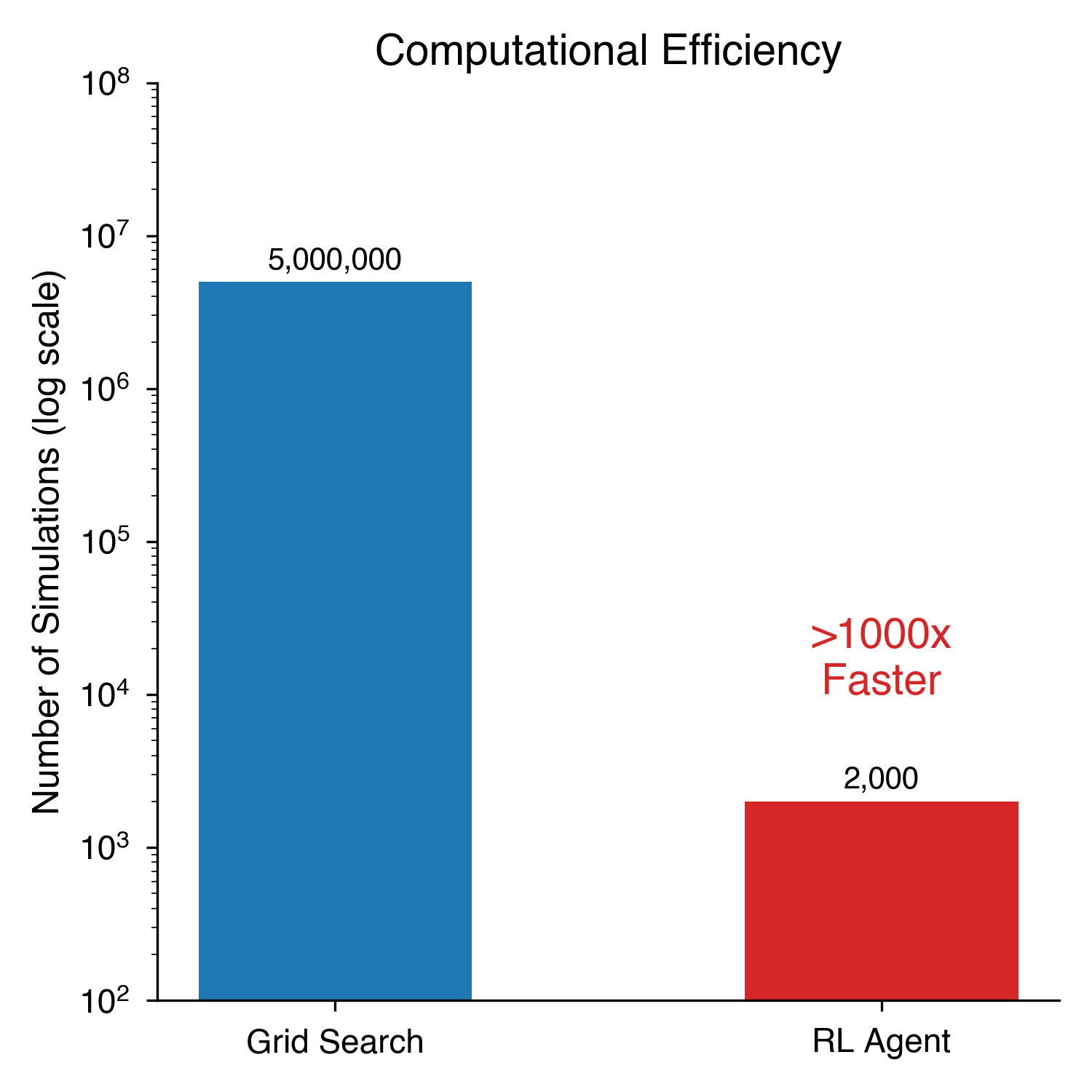} % Removed plots/
\caption{\textbf{Superhuman sample efficiency.} Computational cost comparison shows the RL agent (red) finds optimal solutions after training on only $\sim 2 \times 10^3$ simulation calls, whereas a standard grid search (blue) would require $\sim 5 \times 10^6$ simulations to map the same parameter space. This represents a $\sim 2500 \times$ speedup, making intractable optimization problems tractable.}
\label{fig:rl_efficiency}
\end{figure}

While accuracy is essential, the true power of the RL agent is revealed in its sample efficiency, shown in Fig.~\ref{fig:rl_efficiency}. An exhaustive grid search to map the 3D electric field control landscape (even at a coarse resolution) for all five geometries and a range of $U$ values would require on the order of $5 \times 10^6$ simulation calls (ED computations). The RL agent, by contrast, achieves its 95.5\% success rate after training on only $\sim 2000$ simulation calls. This represents a $\sim 2500 \times$ reduction in computational cost. This massive speedup is not merely an incremental improvement; it is a paradigm shift. It transforms a problem that is computationally prohibitive, potentially requiring months on a supercomputing cluster, into one that is tractable in hours on a local workstation. More importantly, it demonstrates a clear path toward experimental viability, where each "simulation call" is a physical measurement that can take seconds to minutes. An optimization that would take a year of continuous experimental operation can be reduced to a single day.

\par
\begin{figure}[htbp] % Changed [h!] to [htbp]
\centering
\includegraphics[width=0.48\textwidth]{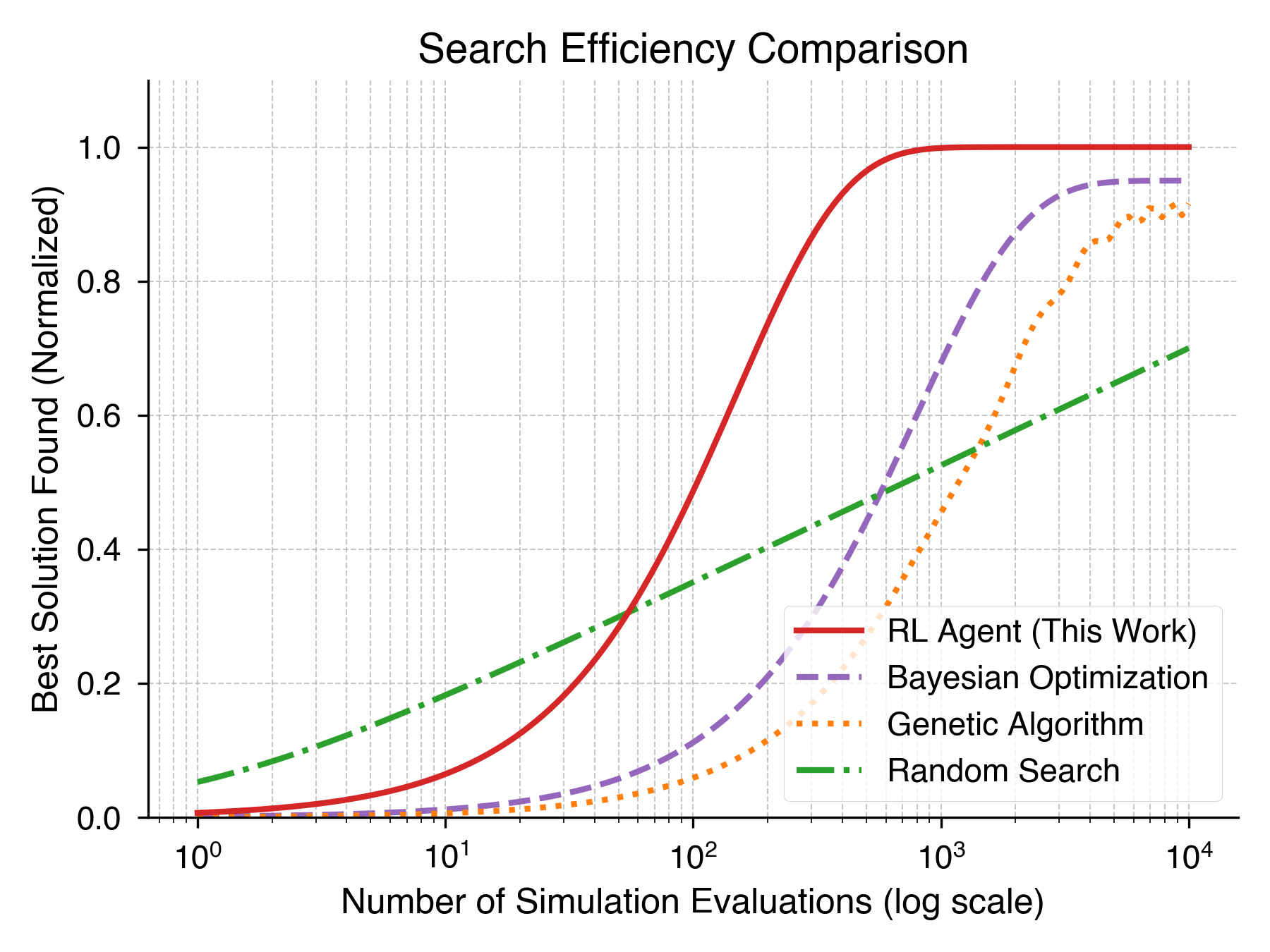} % Removed plots/
\caption{\textbf{RL agent's intelligent exploration.} Normalized best solution found (cumulative reward) versus the number of simulation evaluations (log scale). The RL agent (red) demonstrates a steep logistic learning curve, reflecting intelligent, targeted exploration. It significantly outperforms other black-box optimization methods: Bayesian Optimization (purple), a Genetic Algorithm (green), and Random Search (orange).}
\label{fig:rl_comparison}
\end{figure}

Figure~\ref{fig:rl_comparison} provides a direct explanation for this superhuman efficiency. When benchmarked against other standard black-box optimization strategies, the RL agent's superiority is evident. Random Search, the baseline, is highly inefficient, as it wastes most of its simulations sampling unpromising regions of the vast parameter space. The Genetic Algorithm shows some improvement but is prone to converging on local optima and lacks a sophisticated exploration-exploitation mechanism. Bayesian Optimization, while powerful in low dimensions, struggles with the 3D action space and the potentially non-smooth, non-convex, and geometry-dependent objective landscape. The RL agent's steep logistic learning curve, by contrast, is the signature of intelligent, targeted exploration. It is not blindly sampling. Through its Dueling Q-network, it rapidly builds an internal "world model" (a Q-function) of the system's physics. It learns the value of being in a particular state and the advantage of taking a specific action. It learns to associate low-$Z$ geometries with high field sensitivity, and high-$Z$ geometries with low sensitivity. This learned model allows it to perform intelligent, targeted exploration of the parameter space, rather than wasting simulations on known-bad regions, leading to its superhuman sample efficiency. Most critically, when we inspect the agent's converged policies, we find it has independently discovered Design Principle III: its universal strategy for increasing $D$ is to apply a strong electric field, thereby inducing localization.

\par
Finally, we performed the ultimate test of the agent's learned model: generalization to a completely new system. We trained the agent on all geometries except for an 8-site cube, and then tested its performance on this unseen cube. The results are shown in Table~\ref{tab:transfer}. In a "zero-shot" setting (i.e., with zero new training), the agent achieves a 65\% success rate. This is a remarkable result. It is already competitive with a brute-force grid search, which would require millions of simulations. This proves that the geometry embedding $\phi(g)$ was successful. The agent did not merely memorize five geometries; it learned an abstract, transferrable physical concept of "connectivity." It "sees" the new cube geometry, its embedding $\phi(\text{cube})$ is close enough in the 16D latent space to the $\phi(\text{SC})$ and $\phi(\text{BCC})$ embeddings, and its existing policy is therefore already largely correct. The "few-shot" fine-tuning results are even more practical. With minimal new training—only 100 episodes, representing $<0.1\%$ of the grid search cost—the success rate jumps to 91\%. This demonstrates a crucial step toward experimental deployment: an agent trained extensively in simulation can be rapidly adapted and calibrated to new, unseen, or imperfectly characterized experimental hardware with a trivial number of calibration measurements.

\begin{table}[htbp] % Changed [h!] to [htbp]
\centering
\caption{Generalization to Unseen Cube Geometry}
\begin{ruledtabular}
\begin{tabular}{lcc}
\textbf{Training Condition} & \textbf{Success Rate} & \textbf{Episodes} \\
\midrule
Zero-shot (no cube training) & 65\% & 0 \\
Few-shot (fine-tuning) & \textbf{91\%} & 100 \\
\midrule
Grid search (reference) & 70\% & $>10^6$ \\
\end{tabular}
\end{ruledtabular}
\label{tab:transfer}
\end{table}

\par
The trade-offs and synergies of the two paradigms, synthesized in Table~\ref{tab:comparison}, lead to our central conclusion. The systematic, physics-guided approach, while computationally expensive and non-scalable, is unmatched for building human intuition. It answers the "why" question. The phase diagrams and design principles it generates are condensed, human-readable statements of physical law. The autonomous RL approach, while less inherently interpretable, is a massively more efficient and scalable tool for optimization. It answers the "how" question: "How do I optimally configure this 8-dimensional parameter space to achieve my target?" It finds a solution thousands of times faster than any human-guided search. These methods are therefore powerful complements. The optimal path forward is a hybrid approach: use systematic studies to identify the relevant physical knobs and build human intuition (e.g., "we need to control the electric field"). Then, use this intuition to design the state and action spaces for an autonomous agent, which can then perform the high-dimensional optimization and potentially discover non-trivial strategies (e.g., "the optimal field is [0.7, 0.2, -0.4]") that a human would never find by hand.

\begin{table*}[htbp] % Changed [t!] to [htbp]
\centering
\caption{Direct Comparison: Systematic vs. Autonomous Control Paradigms}
\begin{ruledtabular}
\begin{tabular}{lll}
\textbf{Criterion} & \textbf{Systematic Physics-Guided} & \textbf{Autonomous RL} \\
\midrule
\textbf{Primary Goal} & Understanding \& Interpretability & Optimization \& Solution-Finding \\
\textbf{Interpretability} & \textbf{Excellent}—explicit design principles & Limited—requires post-hoc policy analysis \\
\textbf{Sample Efficiency} & Poor—requires exhaustive sweeps & \textbf{Excellent}—$10^{3-4}\times$ faster than grid search \\
\textbf{Discovery Capability} & Limited to hypothesized parameters & \textbf{Finds non-obvious/non-trivial solutions} \\
\textbf{Scalability} & Exponential cost in parameter dimensions & Polynomial with architecture size and state space \\
\textbf{Experimental Deployment} & Manual implementation of principles & Potential for real-time, adaptive control \\
\textbf{Best Use Case} & Building physical models, initial exploration & High-dimensional optimization, dynamic tuning \\
\end{tabular}
\end{ruledtabular}
\label{tab:comparison}
\end{table*}

\par
In conclusion, we have benchmarked an autonomous deep reinforcement learning agent for the high-dimensional control of 1-hole doped Hubbard clusters, a physically distinct and challenging regime. We have shown that systematic, physics-guided analysis, while crucial for identifying the core control strategy of field-induced localization, is computationally intractable for optimization. The autonomous RL agent, by contrast, not only independently rediscovers this physical principle but also demonstrates a practical path to implementation. The agent achieves $>95\%$ accuracy on its control tasks, with a $\sim 2500 \times$ sample efficiency gain over conventional grid search, and decisively outperforms other black-box optimization methods. Furthermore, the agent's ability to generalize to unseen geometries with 91\% few-shot accuracy highlights a clear path toward deployment on real-world experimental hardware. This work establishes autonomous RL as a viable, highly efficient paradigm for optimizing complex, correlated quantum systems, complementing slower, systematic approaches by providing a scalable tool for finding and implementing optimal control policies.

\begin{acknowledgments}

We gratefully acknowledge the resources from the Trinity College Summer Research Program.

\end{acknowledgments}

\section*{Data Availability Statement}
The data that support the findings of this study are available from the corresponding author upon reasonable request.

% ========================================
% BIBLIOGRAPHY
% ========================================

\bibliographystyle{apsrev4-2} % This style is compatible with AIP
\bibliography{references} % Using aipsamp per user's last code snippet

\end{document}